# Revealing Task Driven Knowledge Worker Behaviors in Open Source Software Communities

Hongrui Wu, Xiaowan Shi & Yutao Ma

## Abstract

Collaborative activities among knowledge workers such as software developers underlie the development of modern society, but the in-depth understanding of their behavioral patterns in open online communities is very challenging. The availability of large volumes of data in open-source software (OSS) repositories (e.g. bug tracking data, emails, and comments) enables us to investigate this issue in a quantitative way. In this paper, we conduct an empirical analysis of online collaborative activities closely related to assure software quality in two well-known OSS communities, namely Eclipse and Mozilla. Our main findings include two aspects: (1) developers exhibit two diametrically opposite behavioral patterns in spatial and temporal scale when they work under two different states (i.e. *normal* and *overload*), and (2) the processing times (including bug fixing times and bug tossing times) follow a stretched exponential distribution instead of the common power law distribution. Our work reveals regular patterns in human dynamics beyond online collaborative activities among skilled developers who work under different task-driven load conditions, and it could be an important supplementary to the current work on human dynamics.

## Introduction

Over the past two decades, we have witnessed the growing effect of software on human society. This trend echoes the famous Marc Andreessen's 2011 proclamation that software is eating the world [1]. Moreover, software-defined anything has been identified by Gartner as one of the top ten strategic technology trends for 2014 [2], implying that we are entering a new era of the information society defined by software.

Compared with proprietary software, open-source software (OSS) brings forth a fundamental shift in the manner in which software is developed and distributed [3]. Due to lower cost, superior security, freedom from vendor lock-in, better quality and other reasons [4], an increasing number of enterprise-level companies have adopted OSS offerings as an essential part of their IT systems [5]. According to the report released by Statista [6], the projected revenue of the global OSS market will reach close to €58 billion in 2020, twice as much as in 2012. The great success of OSS has attracted a lot of attention from scholars in different research fields, such as computer science, management science, and statistical physics.

So far, researchers have investigated why OSS can succeed from various angles and dimensions [7] and found that the success of OSS is, indeed, affected by some determinants, such as innovation model [8], license [9],[10] or intellectual property right [11], social economic factors [12], ideology [13], software quality [14], network embeddedness [15], and language translation [16]. In fact, OSS communities have long been recognized as one of the key factors contributing to the success of many famous OSS projects [8],[17],[18], and previous studies in this area focused mainly on the motivation [19]-[21], participation and contribution [21]-[23] of software developers (also known as knowledge workers), social structure [24]-[28], and member management [29],[30]. However, there has been relatively little research on behavioral aspects of OSS development, which leaves several important questions about developer behavior unanswered [31].

Human behavior has been generally considered to be one of the most significant issues in science since the time of Watson [32]. The timing of many human activities on the Internet, including e-mail communication [33], web browsing [34] and tweeting [35], has been found to share a bursty character with power-law distributed inter-event times (viz. waiting times) [36], enabling us to quantify and understand human dynamics in the virtual world of cyberspace at different scales [37],[38]. Inspired by the work mentioned above, a few researchers in Software Engineering have recently begun to explore the patters and dynamics of some special types of developer behaviors, such as *follow* on GitHub[1] [39], SVN[2] *commit* [40], *co-commit* and e-mail burst [41], and *Q&A* on Stack Overflow[3] [42], to provide insights into how developers work together in different OSS communities. However, a fundamental issue in developer behavior analysis—the dynamics of developers working under different load conditions in an OSS community—is still unclear.

Software development and maintenance of large-scale OSS projects is a knowledge-intensive and labor-intensive process, where developers work independently or collaboratively to perform the tasks assigned to them. For simplicity, in this paper developer workload refers to the number of task assignments received by a developer over a given period of time (e.g. daily, weekly, and monthly), regardless of qualitative aspects of the tasks such as difficulty. Although previous studies in the chronopsychological research field have suggested that, in general, a person's performance in a given task depends directly on his or her functional state [43],[44] and would be reduced with a rapid increase in workload [45],[46], very few large-scale empirical studies on developer behavior were carried out to support such viewpoints. Thus, for this category of knowledge workers on the Internet, the mechanism behind workload's effect on state and behavior, which is the key to solving the above issue, remains unknown up till now.

Unlike those recent studies [39]-[42], our work focuses on developer behavior in *software defect repair* (SDR) (or software bug resolution), which is a basic category of activities closely related to software quality assurance [47]. The SDR activities involve coordination and collaboration among developers[4]. Fortunately, the availability of large volumes of bug-tracking data in OSS projects makes it easier to analyze the patterns and dynamics of developer behaviors (i.e. bug fixing and bug tossing) in the SDR activities. In this paper, we conduct a large-scale experiment on two well-known OSS communities, namely Eclipse[5] and Mozilla[6], and our key finding is that when the working state of developers is transitioned from *normal* into *overload*, they exhibit a different behavioral pattern in which the processing times of both the behaviors, however, do not follow a power-law distribution, which differs from the previous findings on the timing of human behavior in many domains [33]-[38],[40],[52]. Moreover, we propose a phenomenological model to characterize the dynamics of developers dealing with bugs in spatial and temporal scale, possibly leading to an optimal way of bug assignment and developer resource planning and scheduling for OSS projects.

---

[1] https://github.com/
[2] http://subversion.apache.org/
[3] http://stackoverflow.com/
[4] After a bug is verified, it will be assigned to a developer. Then the developer has the option of fixing the bug or tossing (reassigning) the bug to other developers who may be more likely to fix it [48]. In some cases, a bug is tossed among developers many times until it is fixed. For example, on average, it took about 10~60 days to toss out a verified bug in Eclipse [48].

# Results

This section consists of the following three findings. First, by analyzing the relation between workload and performance of developers, we find the evidence of developers' work overload in the SDR activities. Second, we unfold behavioral patterns of developer working in the working states of *normal* and *overload* in temporal scale and spatial scale. Third, we further reveal that the processing times of developer behaviors in our work follow a stretched exponential distribution.

**Perceiving developers' work overload in the SDR activities**

According to the data collected from the SDR activities in the Eclipse and Mozilla communities, we investigated the relation between workload and performance of developers. Here the developer workload is the number of bugs received by a developer in a three-month interval, and the developer performance is measured in terms of fixing rate and tossing rate. More specifically, the average fixing rate of developers for a given workload $N$ is defined as follows:

$$P_f(N) = \frac{\sum_{i=1}^{M} \frac{K_f(i)}{N}}{M}, \qquad (1)$$

where $K_f(i)$ is the number of bugs fixed by developer $i$ and $M$ is the number of developers who assume the same workload. Similarly, the average tossing rate is defined in Eq.(2), where $K_t(i)$ is the number of bugs tossed from developer $i$ to other developers.

$$P_t(N) = \frac{\sum_{i=1}^{M} \frac{K_t(i)}{N}}{M}. \qquad (2)$$

---

**Figure 1: Identification of developers' work overload in Eclipse.**

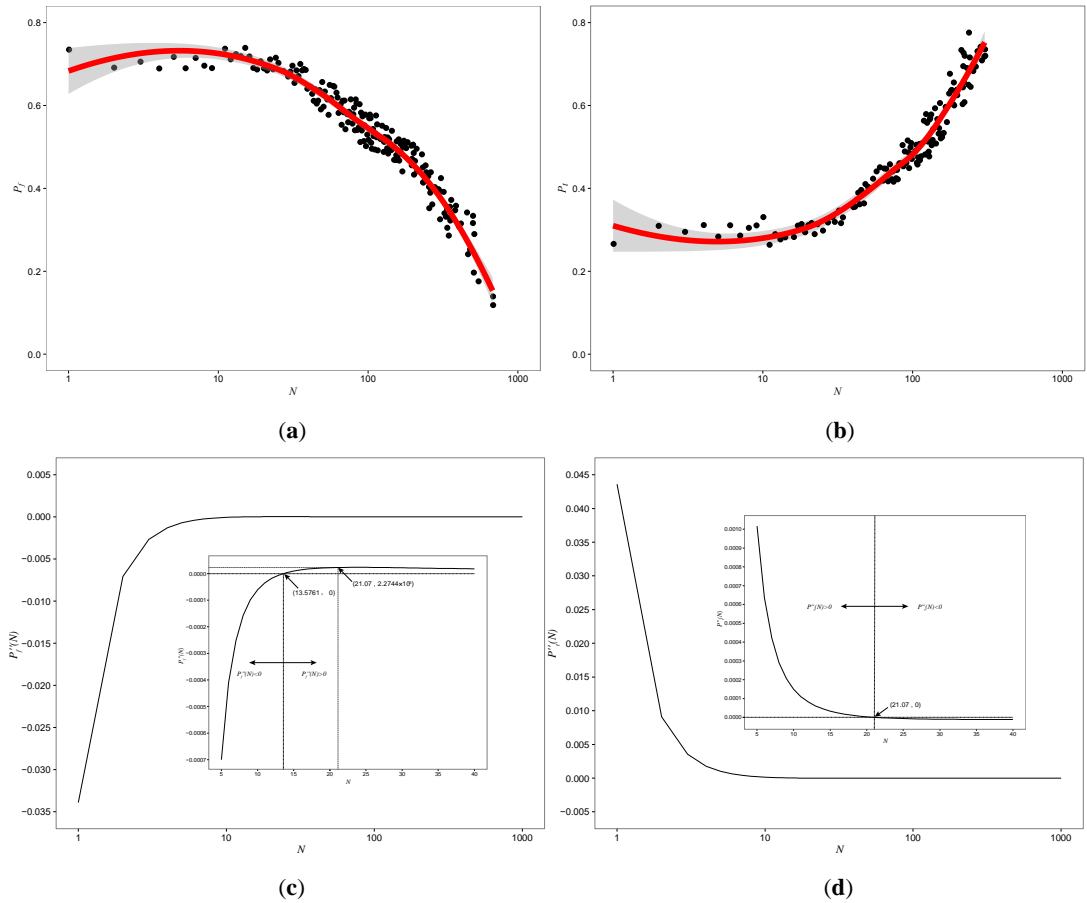

(**a**) The relation between developer workload (*N*) and the average fixing rate of developers ($P_f$), and (**b**) the relation between developer workload and the average tossing rate of developers ($P_t$). Scattered black dots in (**a**) and (**b**) are best fitted by the red curves at a 99% confidence interval (grey shadow), respectively. (**c**) The second derivative of the red fitted curve (for bug fixing) shown in (**a**), and (**d**) the second derivative of the red fitted curve (for bug tossing) shown in (**b**). A more detailed computational procedure of both the second derivatives refers to (**SI**, **Calculation of workload threshold**).

According to the red fitted curves shown in Fig.1, when dealing with an increasing number of bugs, the average fixing rate of developers ($P_f$) is declining and their average tossing rate ($P_t$) is rising. Fig.1(c) and Fig.1(d) further show how the rate of change of developer performance is itself changing with respect to developer workload. It is obvious that the changing rates of the performance of these developers enter into a new stable state (viz. work overload) if they receive, on average, more than 21.07 bugs, i.e., the second derivatives of the functions $P_f(N)$ and $P_t(N)$ are always very small numbers close to zero[7] and their signs remain unchanged when the developer workload (*N*) is greater than 21.07. Theoretically, this fits the requirements of *overload* defined in (Materials and Methods, Defining working states of developers), indicating that in this working state $P_f$ and $P_t$ will maintain steady rates of descent and growth, respectively. Therefore, such a value of *N* is selected as the threshold to determine the transition of developers' working state from *normal* into *overload*. Please note that the results of the Mozilla project refer to (**SI**, **A case study of Mozilla**).

---

[7] In this paper, they are considered to be approximately equal to zero.

## Patterns of developer behaviors in two different working states
### Developers' behavioral patterns in spatial scale

Each developer is assumed to be a bug processing system which manages bug assignments, and the system's working mechanism can be explained using queueing theory. A new bug assignment is inserted at the tail of a queue (also known as a receiving queue), and it will be processed at a specific position in the receiving queue. Developer behaviors of bug processing in spatial scale are formulated by the average first fixing position (AFFP) $Q_f$ and the average first tossing position (AFTP) $Q_t$ in the receiving queue, which are defined in Eq.(3) and Eq.(4), respectively.

$$Q_f = \frac{\sum_{i=1}^{M} S_f(i)}{M}, \tag{3}$$

$$Q_t = \frac{\sum_{i=1}^{M} S_t(i)}{M}, \tag{4}$$

where $S_f(i)$ and $S_t(i)$ are the positions where bugs in the receiving queue are firstly fixed and tossed, respectively, by developer $i$. A smaller $Q_f$ (or $Q_t$) indicates that developers are more inclined to fix (or toss) the bug at the head of the receiving queue, and vice versa for a larger $Q_f$ (or $Q_t$). A simple illustration of these concepts refers to **Fig.S2**

**Figure 2: Eclipse Developers' behavioral patterns in spatial scale.**

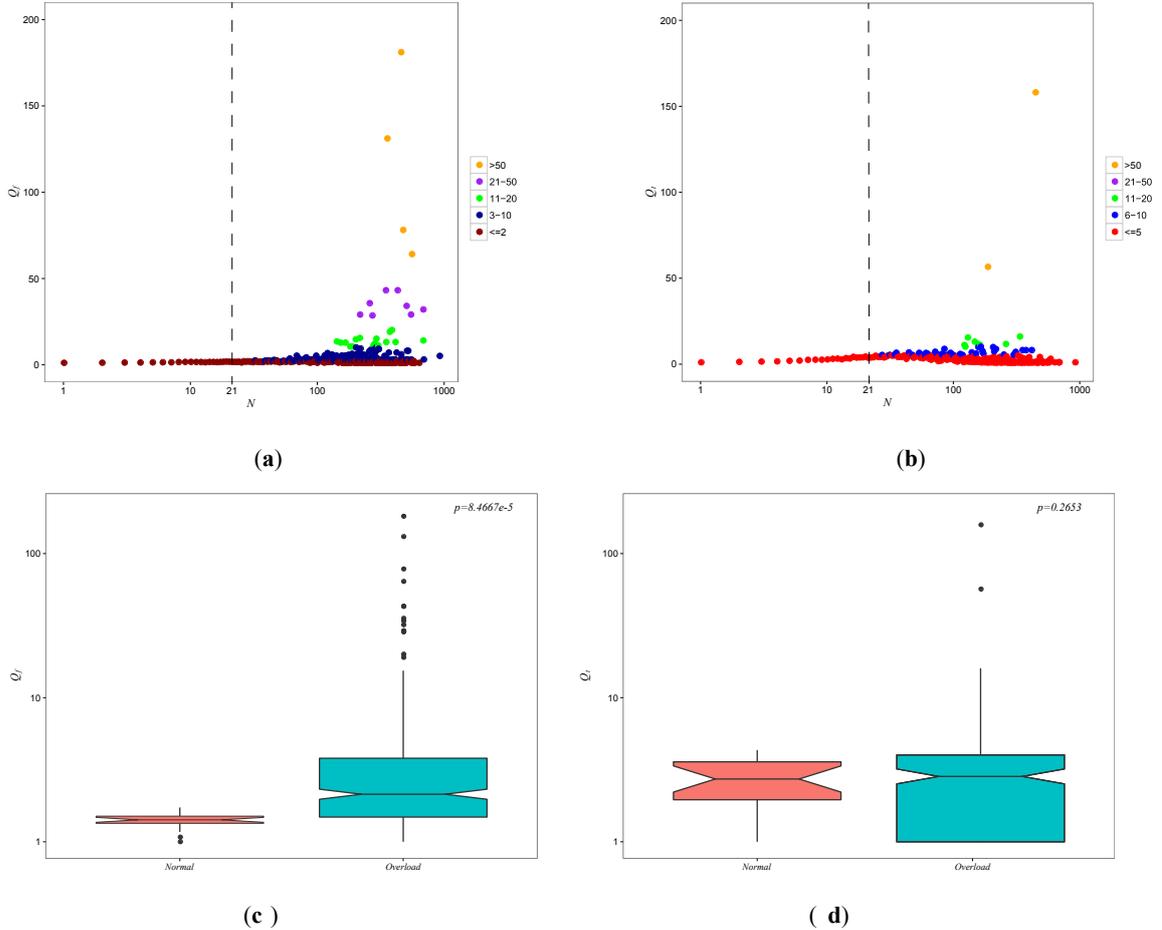

(**a**) Scatter plot of developer workload (*N*, an integer) versus the average first fixing position of developers ($Q_f$),

and (**b**) scatter plot of developer workload versus the average first tossing position of developers ($Q_t$). $Q_f$ and $Q_t$ are grouped by scattered dots with different colors in (**a**) and (**b**), respectively. The vertical dotted lines represent the threshold of the transitions between two different working states. When $N \leq 21$, $Q_f \leq 2$ and $Q_t \leq 5$. (**c**) Box plots of $Q_f$ in the working states *normal* and *overload*, and (**d**) box plots of $Q_t$ in the two working states.

Under the normal working state ($N \leq 21$), the value of AFFP ($Q_f$) is not greater than two, which indicates that developers prefer to fix the bug at the front position in their receiving queues using a common policy of First-In, First-Out (FIFO). However, $Q_f$ increases obviously when the working state is transitioned into *overload* (see Fig.2(a)), implying that a number of developers tend to fix newly received bugs or those ones with high priority first, as well to toss out the oldest bugs to other developers first (see the lower part of the dark green box in Fig.2(d)). According to the result of the Kruskal–Wallis H test performed at the 0.01 level, there is a very significant difference between the two groups of AFFP data in the two working states (see Fig.2(c)). Unlike the finding on AFFP, Fig.2(b) shows that developers' state transitions do not cause obvious changes in the value of AFTP ($Q_t$). Besides, the result of the Kruskal–Wallis H test also indicates that we cannot reject the null hypothesis, which assumes that the two groups of AFTP data are from the same distribution (see Fig.2(d)). That is to say, on average, developers' behavior of bug tossing seems very similar in spatial scale, regardless of their working states. This result suggests that most of developers would try to fix bugs first unless they are assigned too many bugs or can't fix the bugs at hand.

**Developers' behavioral patterns in temporal scale**
**Figure 3: Eclipse Developers' behavioral patterns in temporal scale.**

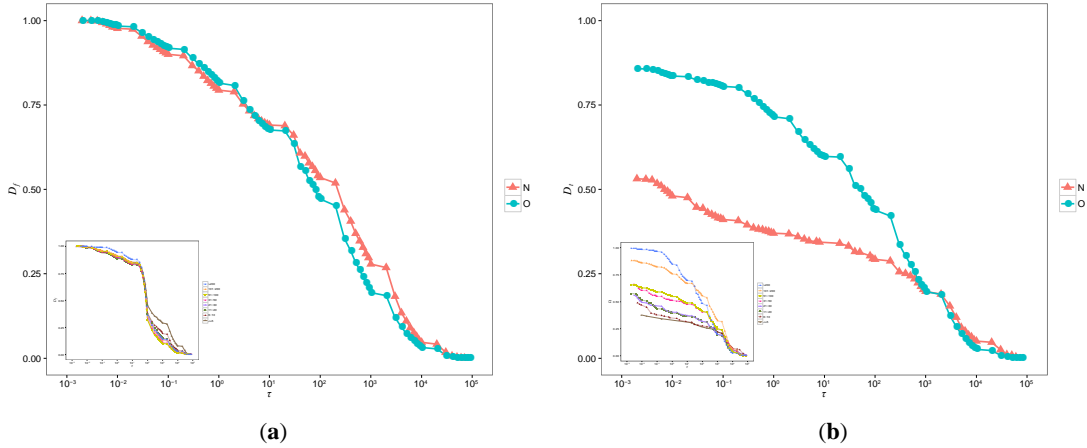

(**a**) $D_f(\tau)$: cumulative developer distribution described in terms of processing time for fixing bugs (in hours), and (**b**) $D_t(\tau)$: cumulative developer distribution described in terms of processing time for tossing bugs (in hours). The cumulative developer distribution function is defined as the probability that a developer chosen at random spends more than a given period of time on processing (fixing or tossing) bugs. N and O denote *normal* and *overload*, respectively. The curved lines with different colors in the two subplots are grouped by developer workload.

During the SDR activities, the processing times of developers usually vary from a few minutes to several months. As shown in Fig.3(a), the red curved line with triangles and the green curved line with circles, which represent the cumulative developer distributions of bug fixing times in *normal* and *overload*, respectively, drop in the same manner. This result indicates that the timings of fixing bugs are strikingly similar in the two working states. Furthermore, the eight curved lines grouped by different workload values in the subplot of Fig.3(a) also follow a similar pattern. Fig.3(b), by contrast, reveals a completely different pattern, which suggests that there is a marked difference between the two cumulative developer distributions of bug tossing times in *normal* and *overload*. This interesting result shows that developers' working states indeed affect their behaviors of bug tossing in temporal scale. As the developer workload increases, an increasing number of developers prefer to toss out excessive bug assignments to others in a short time, e.g. an hour or even less (see the subplot of Fig.3(b)), without careful consideration.

## Distribution of bug processing times

**Figure 4: Distributions of bug fixing times and bug tossing times in Eclipse.**

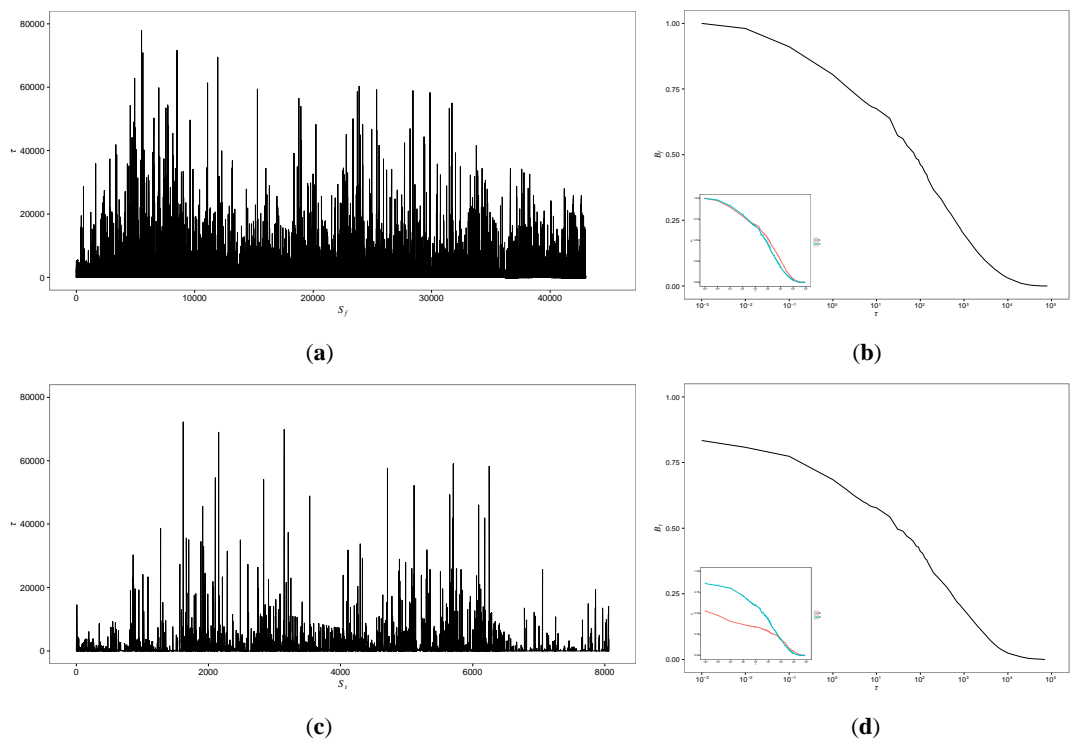

(**a**) Processing times $\tau$ (in hours) of more than 41,000 bug fixing events ($S_f$) sorted by their time of occurrence, (**b**) $B_f(\tau)$: cumulative bug distribution described in terms of processing time (for fixing bugs), (**c**) processing times of more than 8,000 bug tossing events ($S_t$) sorted by their time of occurrence, and (**d**) $B_t(\tau)$: cumulative bug distribution described in terms of processing time (for tossing bugs). The cumulative bug distribution function is defined as the probability that a randomly chosen bug processing event lasts over a given period of time. N and O in the two subplots denote *normal* and *overload*, respectively.

Although about half of bugs in Eclipse were fixed or tossed in 100 hours, a great number of bugs needed more time (e.g. 1,000 hours) to be processed. Because bug fixing normally requires a developer's expertise, skills and experience [49], on average, it usually took 1,237.68 hours to fix a bug, versus 1,163.65 hours for tossing a bug in Eclipse. We find no obvious bursty character in developer behaviors shown in Fig.4(a) and Fig.4(c). Moreover, the cumulative bug distributions of fixing times and tossing times do not follow a typical power-law distribution, which is different from the prior studies on human dynamics [33]-[38],[40],[52]. According to a detailed analysis of curve fitting in (**SI**, **Identifying the distribution of processing times**), the two distributions follow a stretched exponent distribution [50],[51] (also known as the complementary cumulative Weibull distribution), $f_\beta(\tau) \sim e^{-\tau^\beta}$, which has a tail that is "fatter" than the exponential distribution but much less so than a pure power law distribution. Besides that, we also find that the state transitions of developers have very little effect on $B_f(\tau)$ (viz. the timing of bug fixing events) (see the subplot in Fig.4(b)). But, for $B_t(\tau)$, a different pattern of the timing of bug tossing events occurs when the working state of developers is transitioned to *overload* (see the subplot in Fig.4(d)), which is consistent with the result presented in Fig.3(b).

## Discussion

Large-scale online collaborative human dynamics is of great importance in modern human society [52]. The goal of this paper is to investigate online collaborative behavior of knowledge workers from the perspective of software developers. Tasks in OSS communities are often under regulation and control, even without centralized management. First of all, the developer workload related to bug assignments is considered in this paper as a starting point of our work. The phenomenon of work overload is observed during the collaborative SDR activities, and, according to a formal mathematical definition, we identify a workload threshold to determine developers' working state transitions. Moreover, we examine whether such a working state affects behavioral patterns of developers in spatial and temporal scale. Nevertheless, we have to admit that the working states presented here cannot describe developers' true states precisely since we lack sufficient real data of their online and offline behaviors. On the other hand, we are also unable to empirically verify each developer's state at a given time unless we can conduct a large-scale online questionnaire. Even so, we still argue that our analysis sheds light on the underlying mechanisms of complex collaborative activities among developers in a quantitative, comprehensive way.

To the best of our knowledge, few indicators have been proposed to analyze developer behavior in spatial scale. In this paper, the first processing position in a receiving queue, which reflects how a developer prioritizes a sequence of tasks and when he or she processes the higher-priority ones, is utilized to measure developer's decision-making behavior of task processing. Our empirical results show that the ways developers fix bugs are different in two different working states. That is to say, when developers are in the normal working state, they tend to fix bugs using a FIFO queuing policy, but not in the other case. Surprisingly, developers' behaviors of bug tossing are not affected by their working states, according to the Kruskal–Wallis H Test with a high significance level. In addition, AFTP values are greater than those of AFFP under both the two working states. One possible explanation is that developers always give priority to bug fixing due to their sense of duty. But, when faced with excessive tasks, we assume that they may make a tradeoff between task priority and (personal) time availability to maintain a certain level of productivity. An elaborate model that unfolds the basic rules behind the behavioral patterns presented here will be our future work.

The distributions of inter-event times (or waiting times) were widely investigated in the recent decade. Considering the particularity of developer behaviors discussed in this paper, we analyze the processing times of bugs ($\tau$). The empirical results show that, in the working states *normal* and *overload*, the cumulative developer distributions on bug fixing times are almost the same, whereas the cumulative developer distributions on bug tossing times are quite different. When developers are overloaded with tasks, an increasing number of developers tend to toss out the received bugs to others in a relatively short period of time as the workload increases. One possible explanation is that developers would try to alleviate stress at work, while ensuring that those tossed bugs can be fixed by possible developers as soon as possible. Besides, it is noticeable that when $\tau$ is greater than 1,000 hours, the cumulative distributions of developers under the *normal* and *overload* states are very similar (see Fig.3(b) and **Fig.S6(b)**). That is to say, a fraction of developers retain the bugs they received for a very long time and then toss out them, regardless of their working states. For this interesting phenomenon, it may be because the bugs these developers didn't deal with in a timely manner were forgotten, but as time goes by they were found again and then processed.

Taken together, the findings of the above two aspects suggest two distinct behavioral patterns of developers. For the behavior of bug fixing, developers seem to take different queuing strategies when they are in two different working states, probably because the reasonable prioritization of tasks is very helpful in improving the productivity of individual and collective workers [53]-[55], especially once they are overloaded with tasks. Interestingly, the timing of such behavior is not substantially affected by developer's working states. For the behavior of bug tossing, its pattern is completely contrary to that of bug fixing, i.e., developers tend to take similar queuing strategies to toss out excessive tasks irrespective of working states, but the timing of such behavior varies when developers are transitioned to the *overload* state. Unfortunately, the intrinsic relationships between developer's spatial behavior and temporal behavior remain unknown. Thus, our future work is to explore such relationships according to more real data collected from other data sources.

Yet, surprisingly, we find that the processing times in our work (including bug fixing times and bug tossing times) follow a stretched exponential law, though all the previous studies related to timing of human activities focus on inter-event time and find different forms of power-law distributions [33]-[38],[40],[52]. On the one hand, the type of times used in our work is different from those of previous studies, mainly because of the one-shot behavior of bug fixing. On the other hand, due to many advantages such as two adjustable parameters with clear physical interpretation, stretched exponent distributions are a fine complement to power-law distributions [56]. Hence, we believe our work could provide a useful supplement to the existing findings of human dynamics.

Last but not the least, in this paper those software developers we analyzed are actually active and skilled developers who have fixed more than 90% of the total number of bugs recorded in Eclipse and Mozilla for ten years. As the saying goes, able men are always busy. Our finding suggests that able developers also encounter the common problem of work overload, at least in theory, and confirms the fact that worker's states determined based on workload can indeed affect his or her behavior in cyberspace. From the perspective of Management Science, this finding can be applied in OSS projects (or similar online collaborative activities which involves large numbers of knowledge workers) to provide guidelines for task assignment and staff scheduling according to each worker's states. Furthermore, from the perspective of Software Engineering, if we can integrate smart algorithms implementing those guidelines that achieve a dynamic balance between workload and capability into bug tracking systems (or similar artificial systems such as Wikipedia), the group productivity and personal working efficiency would be further improved.

## Materials and Methods

### Data collection

Eclipse is a community for organizations and individuals who develop and use Eclipse-based tools hosted by the Eclipse Foundation, which has over 190 members all over the world. Released under the terms of the Eclipse Public License, the Eclipse projects are free and open-source software, and they are focused on developing an open development platform for building, deploying and managing software applications written in different programming languages. Anyone who has an Eclipse account can report a bug of any Eclipse project to the Eclipse bug database[8] through Bugzilla[9], which is a popular open-source bug tracking system. A new bug report will be assigned to one developer for processing, and the Eclipse bug database stores records of actions taken on a bug within its life cycle [48] (see **Fig.S3**). By using a self-developed software program, we retrieved 200,000 bugs that were fixed and closed during the period from October 11, 2001 to November 2, 2011, and downloaded 850,213 records related to these bugs, which were carried out by more than 5,100 developers, from the Eclipse bug database. Please note that the introduction to the Mozilla project refers to (**SI**, **A case study of Mozilla**).

### Data processing

**Figure 5: A simple example of the modified history of Eclipse bug 146309.**

| Who | When | What | Removed | Added |
|---|---|---|---|---|
| karla.callaghan | 2006-06-14 14:37:38 EDT | Assignee | karla.callaghan | suwanda |
| | | Priority | P3 | P1 |
| | | Target Milestone | --- | 4.3 |
| srinivas.p.doddapaneni | 2006-06-28 11:04:29 EDT | Version | 4.2.1 | 4.2 |
| suwanda | 2006-06-29 09:46:32 EDT | Assignee | suwanda | samwai |
| samwai | 2006-09-11 12:28:44 EDT | Priority | P1 | P3 |
| samwai | 2006-10-30 10:28:32 EST | Target Milestone | 4.3 | 4.4 |
| samwai | 2007-01-31 12:08:45 EST | Target Milestone | 4.4 | future |
| nmehrega | 2007-04-02 18:38:48 EDT | Assignee | samwai | nmehrega |
| | | Summary | Make RAC behaviour the same as on Windows or ask user to decide to proceed | [IAC]Make RAC behaviour the same as on Windows or ask user to decide to proceed |
| | | Target Milestone | future | 4.4i3 |
| nmehrega | 2007-04-03 11:28:32 EDT | Keywords | | plan |
| nmehrega | 2007-04-03 13:40:00 EDT | Keywords | plan | |
| jkubasta | 2007-04-18 07:39:56 EDT | Priority | P3 | P1 |
| nmehrega | 2007-04-30 16:04:53 EDT | Status | NEW | RESOLVED |
| | | Resolution | --- | FIXED |
| suwanda | 2007-05-08 09:49:58 EDT | Status | RESOLVED | CLOSED |
| webmaster | 2016-05-05 10:52:45 EDT | Component | Platform.Communication | TPTP |
| | | Version | 4.2 | unspecified |
| | | Product | TPTP Agent Controller | z_Archived |
| | | Target Milestone | 4.4i3 | --- |

Who: developer ID, When: operation time, What: pre-defined attributes of a bug report, and Removed & Added: updating the content of "What". The definition of a bug's status refers to **Fig.S3**. Because this paper focuses on the behaviors of bug fixing and bug tossing in the SDR activities, we extracted only four records (within red and blue rectangular boxes) from this table for further analysis. It is clear that the original assignee of this bug report is

---

[8] https://bugs.eclipse.org/bugs/
[9] http://www.bugzilla.org/

karla.callaghan. The developer, whose ID is suwanda, received the bug tossed from karla.callaghan on June 14, 2006, and tossed out it to samwai on June 29, 2006. After nearly nine months, this bug was tossed by samwai to nmehrega, and then fixed by the latter on April 30, 2007.

To investigate the timing of bug fixing and bug tossing, we selected 2,794 developers who fixed at least one bug and the related 486,780 records from the raw data we collected. Considering the fact that most of these developers called bug fixers were not active in the SDR activities and fixed a small number of bugs, we only kept the top 20% of fixers in terms of the number of bugs they fixed. In other words, those skilled fixers who had fixed not less than 50 bugs in the period of ten years were considered in our experiment. We then obtained a refined set of data, including 581 developers, 92,779 bugs, and 459,965 records. A simple example of the modified history of a bug is shown in Fig.5. In a statistical sense, 94.49% of records were performed by 20.79% of fixers, which implies that the data which were excluded from the refined data set have very little effect on the main results of this paper. As with many prior studies in Software Engineering [48],[49],[57]-[59], we further filtered out 51,254 bugs without undergoing a full life cycle process, and retained only 148,411 records of bug fixing and bug tossing related to the remaining 41,525 bugs. The server of Bugzilla is located in the eastern US time zone. To avoid the difference between EST (Eastern Standard Time) and EDT (Eastern Daylight Time), they both were converted to GMT (Greenwich Mean Time) for all the records in the refined data set, and the corresponding time values were accurate to the second.

**Defining working states of developers**

For simplicity, the working states of developers are defined by a dynamic relationship between their workloads and performance. If a threshold value of workload is met or exceeded, this will trigger a state transition from *normal* into *overload*. To model such a relationship, the workload is measured by the number of bugs that a developer received during a three-month interval, while the performance of a developer is quantified in terms of fixing rate [60] and tossing rate, which are two measures of how many bugs get fixed and tossed, respectively, among all the bugs that the developer received during the time interval.

Developers have limited capability for processing information, making decisions and performing tasks, and heavy workloads may result in performance issues. Intuitively, a developer tends to toss excessive bugs to other developers when he/she is overloaded with tasks. To find the threshold of workload for developers' state transitions, two mathematical functions $P_f(N)$ and $P_t(N)$, which relate developer workload ($N$) to the mean fixing rate and mean tossing rate of developers, respectively, are obtained by best fitting the corresponding scattered data in our data set with R[10] (see Fig.1 and **Fig.S4**). If $P_f(N)$, $P_t(N)$ and their first derivatives (gradients) are continuous and differentiable, the second derivatives of the two functions, namely the rate of change of gradient, can be calculated, such that we will find their respective inflection points. If the two inflection points are different, we choose the inflection point ($N^*$) which ensures that the following conditions representing a steady state are satisfied as the workload threshold.

---
[10] https://www.r-project.org/

$$\begin{cases} \left|P_f''(N) - c_1\right| \leq \varepsilon, \\ \operatorname{sign}(P_f''(N)) - \operatorname{sign}(P_f''(N*+\varepsilon)) = 0, \\ \left|P_t''(N) - c_2\right| \leq \varepsilon, \\ \operatorname{sign}(P_t''(N)) - \operatorname{sign}(P_t''(N*+\varepsilon)) = 0, \end{cases} \quad (5)$$

where $N$ is greater than $N*$, $c_1$ and $c_2$ are two constants, $\varepsilon$ is a very small real number, and sign() is the sign function in mathematics.

**Distinguishing behavioral patterns in the working states *normal* and *overload***

Our experimental data was divided into two categories by developers' fixing and tossing behaviors in both spatial and temporal scale. On the one hand, developers' behavioral patterns in spatial scale were analyzed in terms of the mean first processing (fixing and tossing) position in their receiving queues of bugs. On the other hand, developers' behavioral patterns in temporal scale were analyzed in terms of the processing time for bugs, which is defined as the duration that a bug is processed by a developer from beginning to end (i.e. resolution or reassignment) (see the examples in Fig.5). For each category of data, we respectively obtained two groups of data samples related to bug fixing and bug tossing under the two working states (see Fig.2, Fig.3, **Fig.S5**, and **Fig.S6**).

Due to different sample sizes, we used a non-parametric method, the Kruskal–Wallis H test [61], to test whether the two groups of data samples in question originate from the same distribution. Based on the null hypothesis that the medians of the two groups to be tested are equal, as well as the alternative hypothesis that their medians are different, the test implemented by MATLAB[11] is executed and then returns a *p*-value. One usually accepts the null hypothesis on condition that the *p*-value is greater than or equal to the value of significance level, e.g. 0.05; otherwise, the null hypothesis is rejected, suggesting that there exists a statistically significant difference between the two groups for testing.

---

[11] http://www.mathworks.com/products/matlab/

## Acknowledgments

The authors thank Haiyang Liu for collecting the raw data of bug reports in Eclipse and Mozilla. The work was supported by the National Basic Research Program of China (973 Program) under Grant No. 2014CB340404, the National Key Research and Development Program of China under Grant No. 2016YFB0800400, the National Natural Science Foundation of China under Grant Nos. 61272111 and 61273216, and the Wuhan Yellow Crane Talents Program for Modern Services Industry.

## Author information

**Hongrun Wu & Xiaowan Shi**
These authors contributed equally to this work.

**Affiliations**
**State Key Laboratory of Software Engineering, Wuhan University, Wuhan, 430072, P.R. China.**
Hongrun Wu, Xiaowan Shi & Yutao Ma

**Contributions**
All authors read the manuscript and provided constructive feedbacks. H.W. performed result analysis and wrote the main manuscript text. X.S. wrote software programs and analyzed the data. Y.M. designed the experiments and finalized the manuscript.

**Competing interests**
The authors declare no competing financial interests.

**Corresponding author**
Correspondence to Yutao Ma (ytma@whu.edu.cn).

# Supplementary Information

**Calculation of workload threshold**

First, by using R, we obtained the best fitting curves for the scattered black dots shown in **Fig.1(a)** and **Fig.1(b)**, respectively, at a 99% confidence interval. Their equations are described as follows.

$$P_f(N) = \frac{1009 N^{\frac{301}{1250\ln 10}}}{1000} - \frac{3041 N^{\frac{791}{1250\ln 10}}}{10000}, \tag{S1}$$

$$P_t(N) = \frac{2019 N^{\frac{141}{100\ln 10}}}{100000} + \frac{1443}{5000 N^{\frac{421}{1250\ln 10}}}, \tag{S2}$$

where $N$ is the developer workload ($N \geq 1$).

Second, since $P_f(N)$ and $P_t(N)$ are continuous and differentiable functions, we obtained their respective first derivatives with respect to $N$, as shown in Fig.S1. Moreover, their equations are described below.

$$P_f'(N) = \frac{303709 N^{\frac{301}{1250\ln 10}-1}}{1250000 \ln 10} - \frac{2405431 N^{\frac{791}{1250\ln 10}-1}}{12500000 \ln 10}, \tag{S3}$$

$$P_t'(N) = \frac{284679 N^{\frac{141}{100\ln 10}-1}}{10000000 \ln 10} - \frac{607503}{6250000 (\ln 10) N^{\frac{421}{1250\ln 10}+1}}. \tag{S4}$$

**Figure S1: First derivatives of $P_f(N)$ and $P_t(N)$ with respect to $N$**

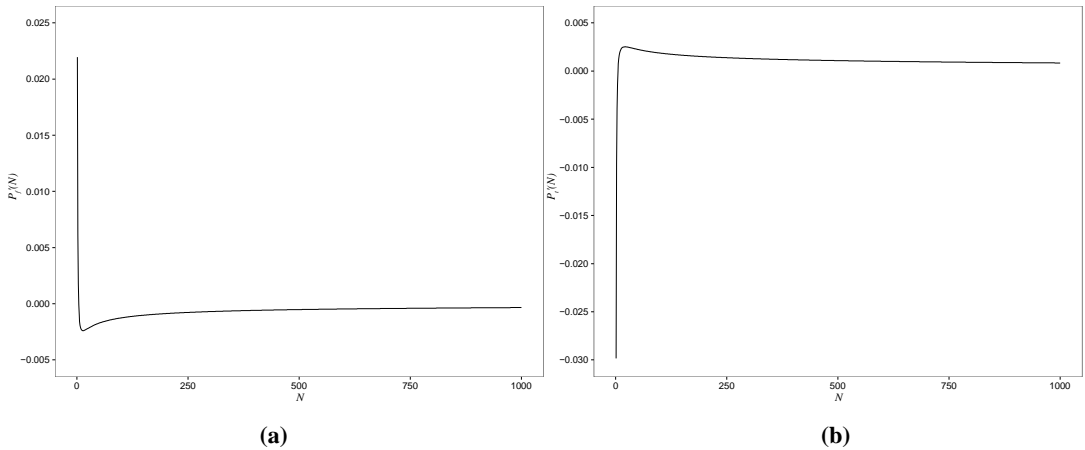

(a) (b)

(**a**) The first derivative of $P_f(N)$, and (**b**) the first derivative of $P_t(N)$. The value of $P_f'(N)$, the rate of change of the mean fixing rate, declines sharply with increasing workload, and then goes up slowly and gradually approaches zero. But, $P_t'(N)$, compared with $P_f'(N)$, shows an opposite trend in value.

Third, according to Eq.(S3) and Eq.(S4), we further obtained the respective derivatives of $P_f'(N)$ and $P_t'(N)$, namely the second derivatives of $P_f(N)$ and $P_t(N)$ (see **Fig.1(c)** and **Fig.1(d)**), as formulated below.

$$P_f''(N) = \frac{303709\left(\frac{301}{1250\ln 10}-1\right)N^{\frac{301}{1250\ln 10}-2}}{1250000\ln 10} - \frac{2405431\left(\frac{791}{1250\ln 10}-1\right)N^{\frac{791}{1250\ln 10}-2}}{12500000\ln 10}, \quad (S5)$$

$$P_t''(N) = \frac{284679\left(\frac{141}{100\ln 10}-1\right)N^{\frac{141}{100\ln 10}-2}}{10000000\ln 10} + \frac{607503\left(\frac{421}{1250\ln 10}+1\right)}{6250000(\ln 10)N^{\frac{421}{1250\ln 10}+2}}. \quad (S6)$$

Finally, let the two second derivatives equal to zero, we obtained two different inflection points (see **Fig.1(c)** and **Fig.1(d)**), namely 13.58 for $P_f(N)$ and 21.07 for $P_t(N)$. Note that, for **Eq.(5)**, $c_1$ and $c_2$ were set to zero, and $\varepsilon$ was set to 0.5e-5. According to the definition of the workload threshold ($N^*$), in this paper $N^*$ was set to 21.07.

## An illustration of receiving queue and related concepts

Each developer will receive a number of bugs, and he/she may fix or toss out some of the received bugs. To simulate developer's decision-making mechanism of task processing, we assume all the received bugs are placed in a queue ordered by receiving time, which is called "receiving queue". An example of such a receiving queue of a specific developer is shown in Fig.S2. In the receiving queue, $B_3$, $B_4$, and $B_6$ (displayed in orange) are tossed to other developers, and the other four bugs (displayed in dark green) are fixed by the developer. For the developer, the first fixing position is one (for $B_1$) and the first tossing position is three (for $B_3$).

**Figure S2: An illustration of receiving queue and related concepts**

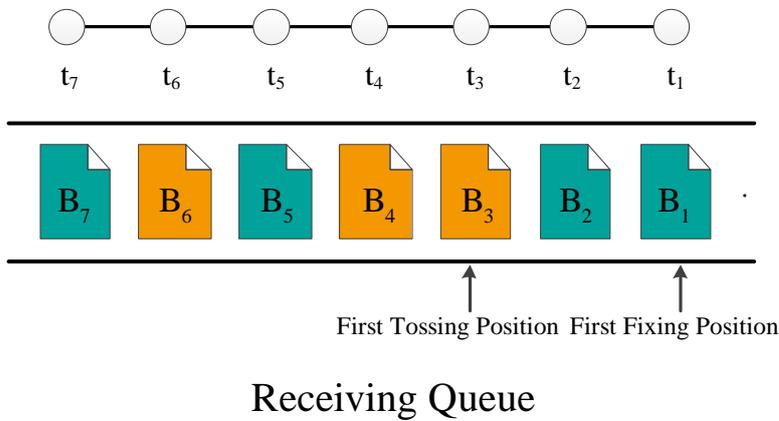

Each bug $B_i$ ($i = 1\sim7$) has a timestamp ($t_i$) of the request when it is received by a given developer, and we sort the bugs by their timestamps in ascending order ($t_1 < t_2 < \ldots < t_7$). Each dark green document icon represents a bug that is fixed by the developer, while each orange document icon denotes a bug that is tossed out to other developers.

## Identifying the distribution of processing times

In our work, the distributions of processing times (see **Fig.4(b)** and **Fig.4(d)**) are determined by the popular quantitative method inspired by Clauset *et al.* [62,63]. To identify the best fitting curve for our empirical data, we compared five commonly used statistical distributions, i.e. power-law ("pl" for short), exponential ("expn"), stretched exponential ("stexp"), log-normal ("lgn"), and power-law with exponential cutoff ("pl-cut"), using the Vuong closeness test [64]. According to the results shown in Tab.S1 and Tab.S2, it is obvious that the stretched exponent distribution ("stexp") is the best among them. In particular, the values for the parameter $\beta$ of the stretched exponent distribution are 0.3299 and 0.2709 for bug fixing and bug tossing, respectively.

**Table S1: A comparison of the five statistical distributions for bug fixing times in Eclipse.**

A cell in this table represents the log-likelihood ratio of one distribution in the row of the cell (or on the left hand side) versus another one in the corresponding column of the cell (or on the right hand side). If the value of a cell is negative, this suggests that the left-hand distribution is closer to the empirical data; otherwise, the right-hand distribution is preferred. Note that *p*-values of all the Vuong closeness tests are equal to zero, indicating that we can reject the null hypothesis that the two distributions are equally close to the true data.

|        | pl    | expn  | stexp | lgn   | pl-cut |
|--------|-------|-------|-------|-------|--------|
| pl     | /     | -73.2 | 16.7  | 12.7  | 11.6   |
| expn   | 73.2  | /     | 73.3  | 73.3  | 73.3   |
| stexp  | **-16.7** | **-73.3** | /     | **-97.2** | **-73.3** |
| lgn    | -12.7 | -73.3 | 97.2  | /     | 24.4   |
| pl-cut | -11.6 | -73.3 | 73.3  | -24.4 | /      |

**Table S2: A comparison of the five statistical distributions for bug tossing times in Eclipse.**

|        | Pl    | expn  | stexp | lgn   | pl-cut |
|--------|-------|-------|-------|-------|--------|
| pl     | /     | -31.8 | 7.24  | -5.5  | 4.9    |
| expn   | 31.8  | /     | 31.8  | 31.8  | 31.8   |
| stexp  | **-7.24** | **-31.8** | /     | **-41.7** | **-1.97** |
| lgn    | 5.5   | -31.8 | 41.7  | /     | 10.3   |
| pl-cut | -4.9  | -31.8 | 1.97  | -10.3 | /      |

## The life cycle of bugs in open-source software

Once a developer (or user) finds a software bug, he/she reports the bug to a bug tracking system (e.g. Bugzilla). A bug report is then formed in the system, and it often has a specific status, e.g. "new", "assigned", or "verified", which changes according to the current processing result of the bug. This bug is assigned to a developer after it is confirmed as a new one; otherwise, it is marked as "resolved". After a developer receives the bug, this developer spends a few hours or a couple of days (or months) on trying to fix it. Sometimes, the developer may toss out the bug to other developers if he/she cannot fix it. In this process, if the bug is found to be "invalid", "duplicate", "fixed", "worksforme", or "wontfix", it will be marked as "resolved". For any bug marked as "resolved", if its resolution is "FIXED", the bug is marked as "verified" and will then be closed. Of course, this bug may be reopened on occasion because of new problems.

**Figure S3: The life cycle of a bug report [48]**

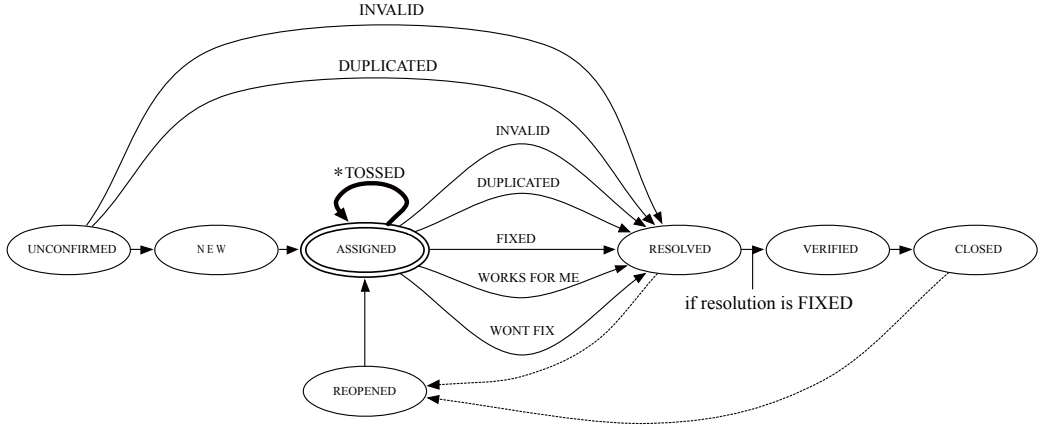

## A case study of Mozilla

Mozilla is also a free-software community created in 1998 by the members of Netscape. The community is supported institutionally by the Mozilla Foundation and Mozilla Corporation, and has over 40,000 active contributors from across the globe. Mozilla has developed a few famous products such as Firefox, Thunderbird, and Bugzilla. Similarly, we retrieved 220,000 bugs that were fixed and closed during the period from October 11, 2001 to November 2, 2011, and downloaded 894,990 records related to these bugs, which were carried out by more than 9,000 developers, from the Mozilla bug database. According to the above-mentioned method of data processing in (**Materials and Methods**, **Data Processing**), we obtained a refined data set comprises of 1,045 skilled fixers, 60,673 bugs with full lifecycle management, and 226,655 records related to these bugs.

By using the method presented in (**Materials and Methods**, **Defining working states of developers**), we obtained the best fitting curves for the empirical data of the Mozilla community (see Fig.S4(a) and Fig.S4(b)) and their respective first and second order derivatives (see Fig.S4(c) ~ Fig.S4(f)). Their equations are described as follows.

$$P_f(N) = \frac{7241 N^{\frac{4059}{100000}}}{10000} - \frac{2739 N^{\frac{4699}{10000}}}{100000}, \tag{S7}$$

$$P_f'(N) = \frac{29391219}{1000000000 N^{\frac{95941}{100000}}} - \frac{12870561}{1000000000 N^{\frac{5301}{10000}}}, \tag{S8}$$

$$P_f''(N) = \frac{68226843861}{10000000000000 N^{\frac{15301}{10000}}} - \frac{2819822942079}{100000000000000 N^{\frac{195941}{100000}}}, \tag{S9}$$

$$P_t(N) = \frac{113 N^{\frac{569}{1000}}}{10000} + \frac{2901}{10000 N^{\frac{3469}{50000}}}, \tag{S10}$$

$$P'_t(N) = \frac{64297}{10000000 N^{\frac{431}{1000}}} - \frac{10063569}{500000000 N^{\frac{53469}{50000}}}, \quad \text{(S11)}$$

$$P''_t(N) = \frac{538088970861}{25000000000000 N^{\frac{103469}{50000}}} - \frac{27712007}{10000000000 N^{\frac{1431}{1000}}}. \quad \text{(S12)}$$

**Figure S4: Identification of developers' work overload in Mozilla.**

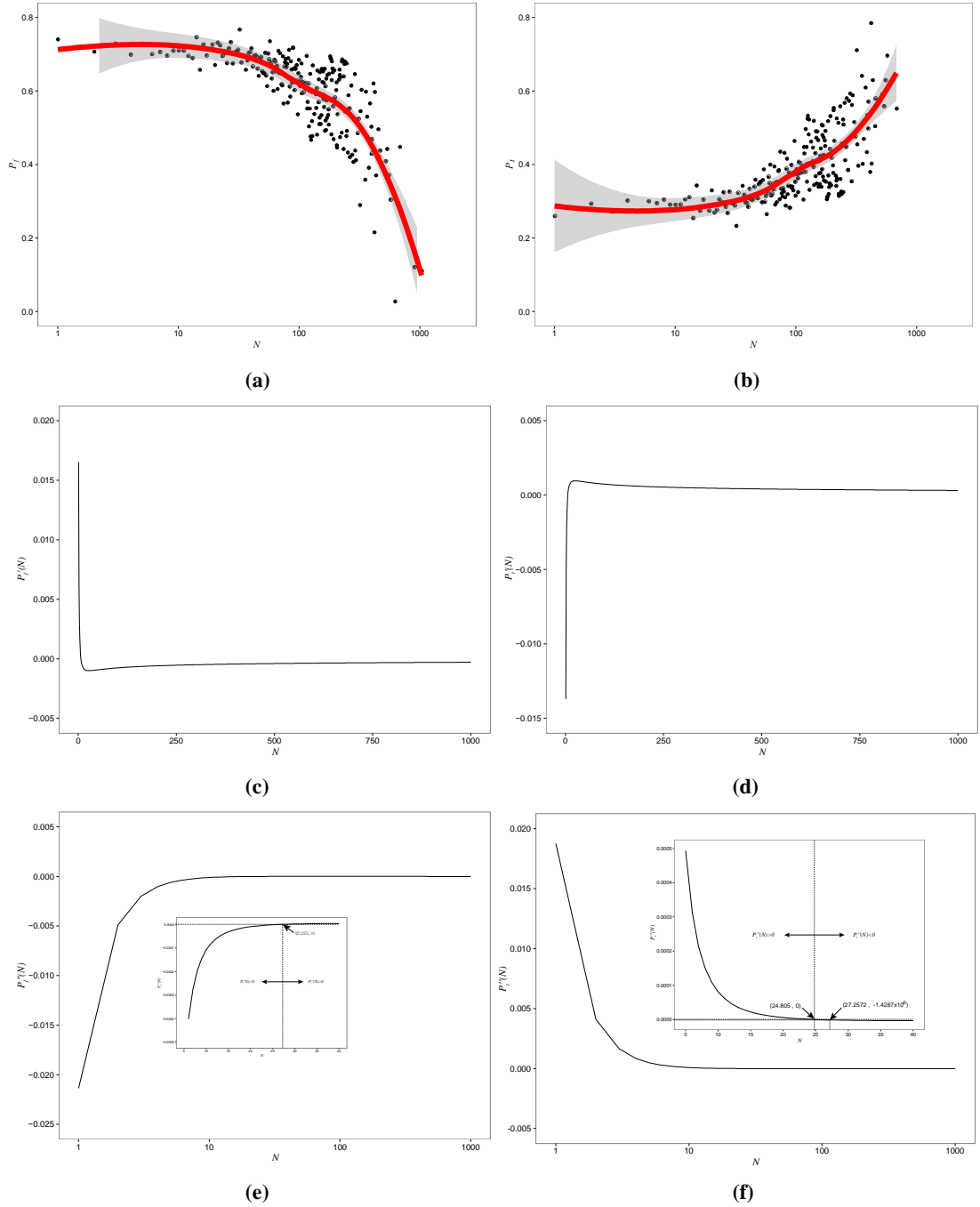

(**a**) The best fitting curve of $P_f(N)$ (at a 99% confidence interval), (**c**) the first derivative of $P_f(N)$, and (**e**) the second derivative of $P_f(N)$. (**b**), (**d**), and (**e**) are the best fitting curve, the first derivative, and the second derivative, respectively, of $P_t(N)$. According to the definition of the workload threshold, $N^*$ is approximately equal to 27.26.

**Figure S5: Mozilla Developers' behavioral patterns in spatial scale.**

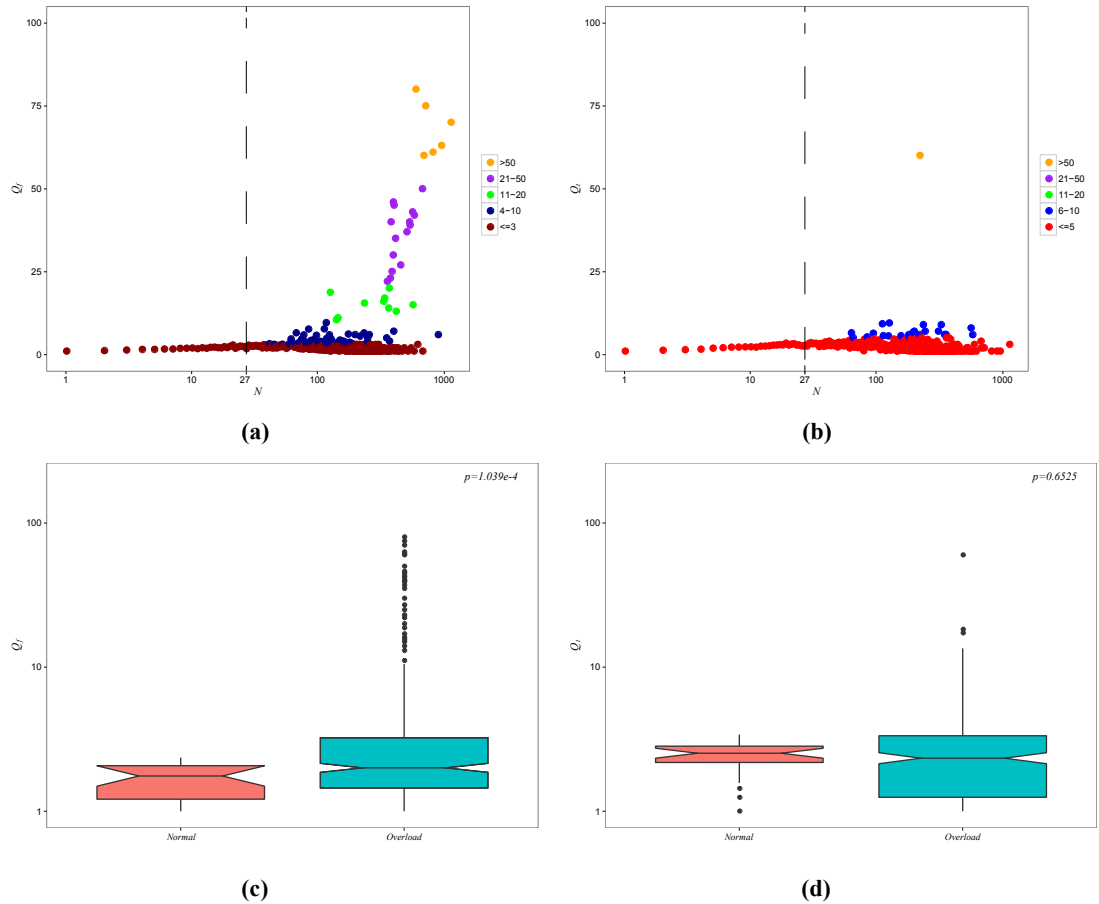

(a)  (b)  (c)  (d)

The meaning of each plot in this figure is the same as that of each plot in **Fig.2**. According to the results of the Kruskal–Wallis H test performed at the 0.01 level, we find the same outcome as shown in the Eclipse community.

**Figure S6: Mozilla Developers' behavioral patterns in temporal scale.**

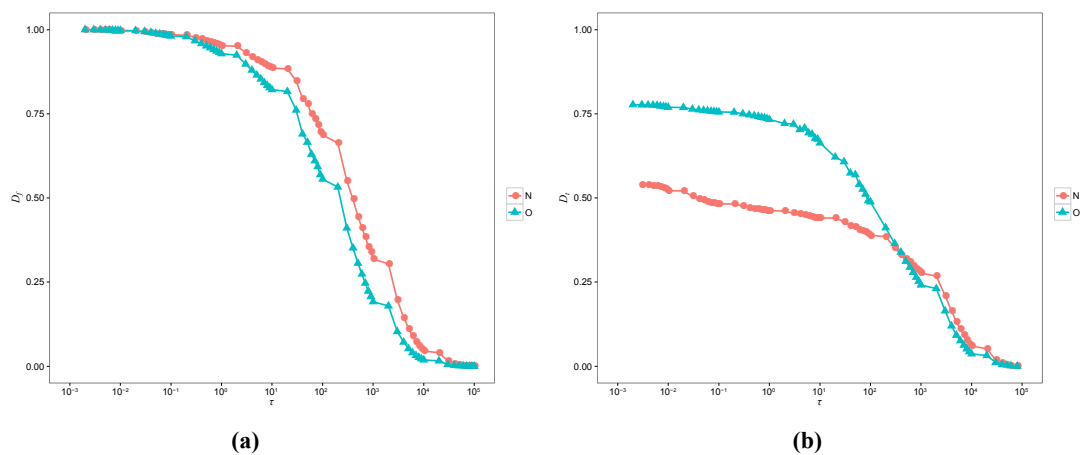

(a)  (b)

The meaning of each plot in this figure refers to **Fig.3**. Compared with **Fig.3**, this figure does not contain the subplots that display the cumulative developer distributions with different developer workloads. Although Mozilla is bigger than Eclipse and consists of different developers, the outcomes of developers' behavioral patterns in temporal scale are almost the same in these two communities.

**Figure S7: Distributions of bug fixing times and bug tossing times in Mozilla.**

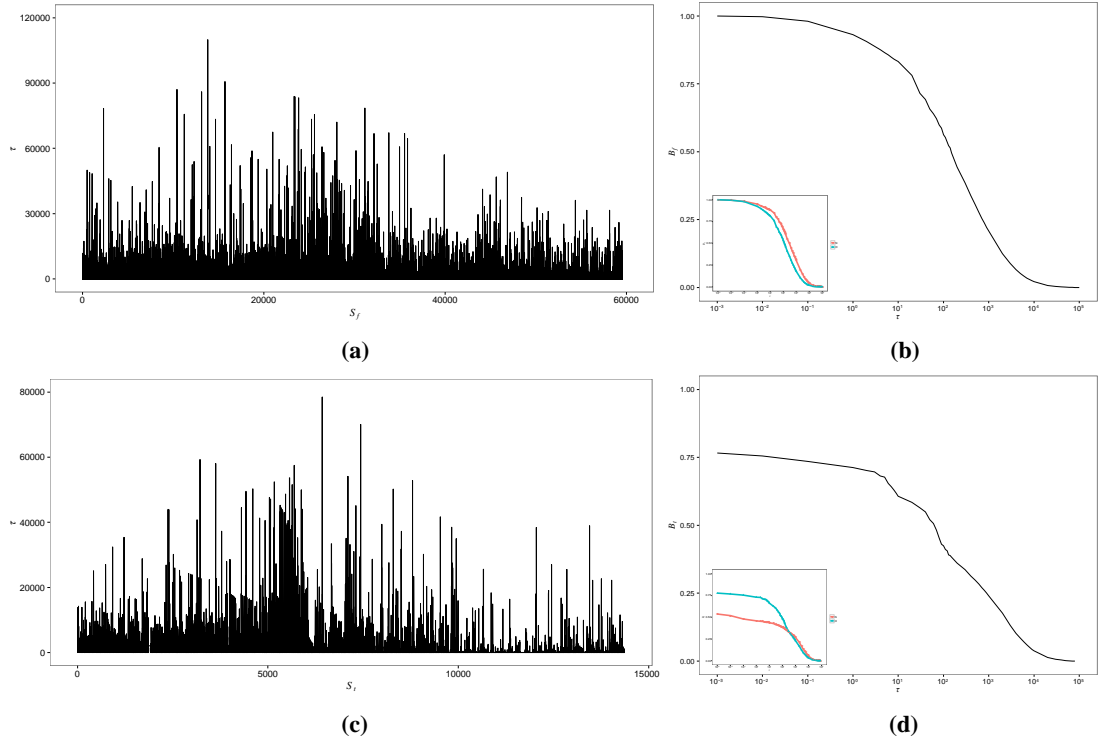

(a)

(b)

(c)

(d)

(**a**) and (**c**) are processing times of bug fixing and bug tossing, respectively. (**b**) and (**d**) are cumulative bug distributions described in terms of bug fixing times and bug tossing times, respectively. According to the results of Tab.S3 and Tab.S4, such distributions of processing times follow a stretched exponential law, i.e. $e^{-\tau^{\beta}}$ (the values of $\beta$ are 0.4500 and 0.4293 for bug fixing and bug tossing, respectively).

**Table.S3: A comparison of the five statistical distributions for bug fixing times in Mozilla.**

|        | Pl    | expn  | stexp | lgn   | pl-cut |
|--------|-------|-------|-------|-------|--------|
| pl     | /     | -36.7 | 2.9   | 17.9  | -5.9   |
| expn   | 36.7  | /     | 36.7  | 36.7  | 36.7   |
| stexp  | **-2.9** | **-36.7** | /  | **-13.7** | **-21.3** |
| lgn    | -17.9 | -36.7 | 13.7  | /     | -30.2  |
| pl-cut | 5.9   | -36.7 | 21.3  | 30.2  | /      |

**Table.S4: A comparison of the five statistical distributions for bug tossing times in Mozilla.**

|        | Pl     | expn  | stexp | lgn    | pl-cut |
|--------|--------|-------|-------|--------|--------|
| pl     | /      | -14.0 | 33.2  | 5.9    | 10.5   |
| expn   | 14.0   | /     | 14.0  | 14.0   | 14.0   |
| stexp  | **-33.2** | **-14.0** | / | **-37.1** | **-7.0** |
| lgn    | -5.9   | -14.0 | 37.1  | /      | 6.1    |
| pl-cut | -10.5  | -14.0 | 7.0   | -6.1   | /      |